\documentclass[12pt,preprint]{aastex}

\newcommand{\be}{\begin{equation}}
\newcommand{\ee}{\end{equation}}

\begin{document}

\title{Intermittence of the Map of Kinetic Sunyaev-Zel'dovich Effect and
    Turbulence of IGM}

\author{Weishan Zhu$^{1,2}$, Long-long Feng$^{1}$ and Li-Zhi Fang$^{2}$}

\altaffiltext{1}{Purple Mountain Observatory, Nanjing, 210008, China}
\altaffiltext{2}{Department of Physics, University of Arizona, Tucson, AZ 85721}

\begin{abstract}

We investigate the possibility of detecting the turbulent state of the IGM with
the kinetic Sunyaev-Zel'dovich (kSZ) effect. Being sensitive to the
divergence-free component of the momentum field of the IGM, the kSZ
effect might be used to probe the vorticity of the turbulent
IGM. With cosmological hydrodynamical simulation in the concordance
$\Lambda$CDM universe, we find that the structure functions of 2D
kSZ maps show strong intermittence, and the intermittent exponents follow
a law similar to the She-Leveque scaling formula of fully developed turbulence. We
also find that the intermittence is weak in the maps of thermal
Sunyaev-Zel'dovich (tSZ) effect. Nevertheless, the superposition of
the kSZ and tSZ effects still contain significant intermittence. We
conclude that the turbulent behavior of the IGM may be revealed by
 the observation of SZ effect on angular scales equal to or less than
0.5 arcminute, corresponding to the multipole parameter $l\geq 2
\times10^4$.

\end{abstract}


\section{Introduction}

The nonlinearly evolved cosmic baryon fluid is probably in the
state of fully developed turbulence in the scale free range. With
hydrodynamical simulation in the concordance $\Lambda$CDM model,
the scaling exponents of the velocity structure functions 
of the intergalactic medium(IGM) are found to follow the She-Leveque 
scaling law(She \& Leveque 1994) from $ \sim 10 h^{-1}$Mpc down to the
dissipation scale (He et al. 2006; Fang \& Zhu, 2011). The density
field of the IGM is in good agreement with the
log-Poisson hierarchy model(Liu \& Fang 2008), which characterizes
the statistical features of fully developed turbulence (Dubrulle
1994; She \& Waymire 1995; Benzi et al. 1996). The intermittent
behavior of the transmitted flux of moderate and high redshifts QSO's
Ly$\alpha$ absorption can also be well explained with the log-Poisson
hierarchy cascade (Lu et al. 2009,2010).

Recently, it is further found that the IGM flow is not potential,
but consists of vorticity, $\omega=\nabla\times {\bf v}$, on scales from hundreds kpc
to a few Mpc  at $z\sim 0$(Zhu et al. 2010). The power spectra of the
vorticity field $P_{\omega}(k)$ and the velocity field $P_{v}$
satisfy the relation $P_{\omega}(k)=k^2P_{v}(k)$ on scales from
$0.2$ to about $3 h^{-1}$ Mpc, indicating that the
velocity field of the IGM is dominated by vorticity on this scale
range.

These results motivate us to study the imprints of the turbulent IGM
in the kinetic Sunyaev-Zel'dovich (kSZ) effect. The kSZ effect is the CMB
temperature fluctuations due to the scattering of CMB photon
by the bulk motion of free electrons in the IGM with
radial peculiar velocity field ${\bf v}(t,{\bf r})$ (Sunyaev
\& Zel'dovich  1972,1980). It can be given by an integral
along a line of sight $s$ as
\begin{equation}
\Delta T(\hat{\bf n})/T=b(\hat{\bf n})= - \int \sigma_T
\left (\frac{n_e{\bf v}\cdot \hat{\bf n}}{c}\right )ds
\end{equation}
where $\sigma_T$ is the Thompson scattering cross section. The quantity $n_e$ is
the number density of electrons, and ${\bf n}$ is the unit vector in
the direction of the line of sight. It has been pointed out that the
kSZ effect is largely contributed by the divergence-free
component of the momentum density field $n_e{\bf v}$. The kSZ effect
would grow fast when the velocity field transits from a
gradient, or curl-free field to curl-dominant one (Vishiniac
1987; Ma et al. 2002; Zhang et al 2004). In other words, the kSZ
effect is sensitive to the vorticity of the IGM velocity field.

Many studies on the kSZ effect have been done with semi-analytical method
or hydrodynamical simulation (e.g. Springel et al. 2001;  da Silva
et al. 2001;  Zhang et al. 2002, 2004; Roncarelli et
al. 2007; Atrio-Barandela et al 2006, 2008; Cunnama et al 2009).
Most of these works mainly focused on the one-point probability density
function distributions, the power spectrum, and their difference with
the corresponding results of the tSZ effect.
The tSZ effect, usually in terms of $y({\bf n})$, is calculated by a line
 integral of the pressure of the IGM (Sunyaev \& Zeldovich, 1980). In this
letter, we use structure functions to extract the non-Gaussian
features of two-dimensional kSZ and tSZ maps composed from hydrodynamical
simulation in the $\Lambda$CDM framework. We find that the
kSZ maps are highly intermittent, and the intermittent exponents are
consistent with model of fully developed turbulence.  While for the tSZ maps
the intermittence is relatively weak. For the maps given by the
superposition of the kSZ and tSZ effects, the intermittent
features are still significant on angular scale $l\sim 2 \times
10^4$. The kSZ effect provides a useful tool to probe the
turbulent state of the IGM.

\section{Map making and power spectrum}

{\it 1. Cosmological simulation and  map making.}  We perform our
cosmological simulation using the WIGEON code (Feng et al. 2004),
which is a hybrid cosmological hydrodynamic/N-body code based on weighted essentially
non-oscillatory (WENO) method. It has been used to study the tSZ effect(Cao et al. 2007),
Ly$\alpha$ forests (Lu et al 2009, 2010), the vorticity of the IGM
(Zhu et al 2010) and the effect of IGM turbulence on the baryon fraction (Zhu
et al. 2011). The simulation is evolved from $z=99$ to $z=0$ in a periodic box
of side length 25 $h^{-1}$
Mpc with a $512^{3}$ grid and an equal number of dark matter particles.
The box size 25 $h^{-1}$ Mpc is reasonable, as it is much larger than
the typical scales of vortices. To test the convergence, we also run a comparison
simulation in a larger box of 100 $h^{-1}$ Mpc with the number of grid
and particles remain $512^3$. The cosmological parameters adopted here are
$(\Omega_{m}, \Omega_{\Lambda},h,\sigma_{8},\Omega_{b},n_{s})=
(0.274,0.726,0.705,0.812,0.0456,0.96)$ (Komatsu et al., 2009).
Radiative cooling and heating are the same as Theuns et al. (1998)
with a primordial composition of $X=0.76, Y=0.24$, and star formation
and its feedback are not included.

\begin{figure}[htb]
\begin{center}
\includegraphics[width=8.0cm]{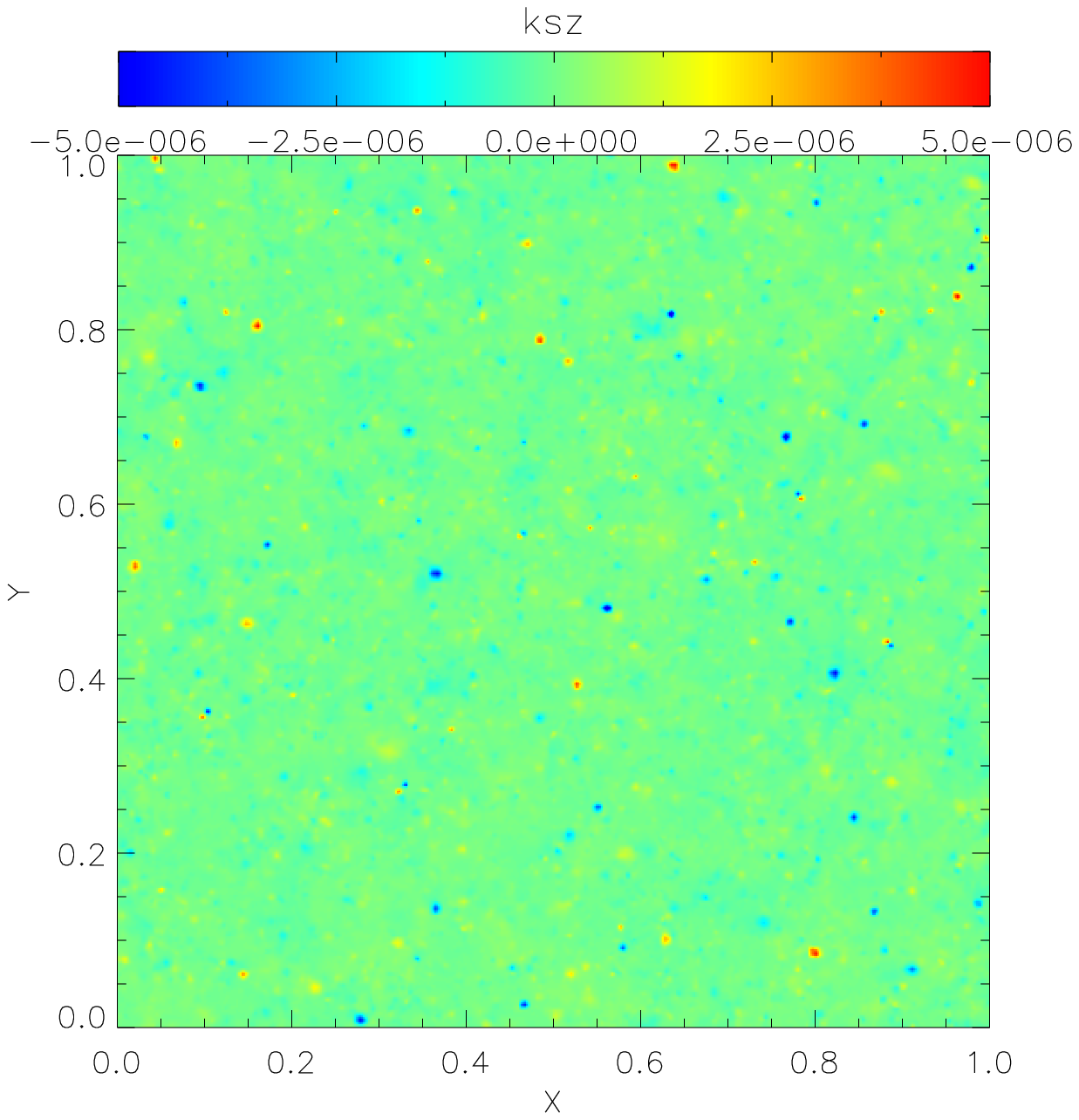}
\includegraphics[width=8.0cm]{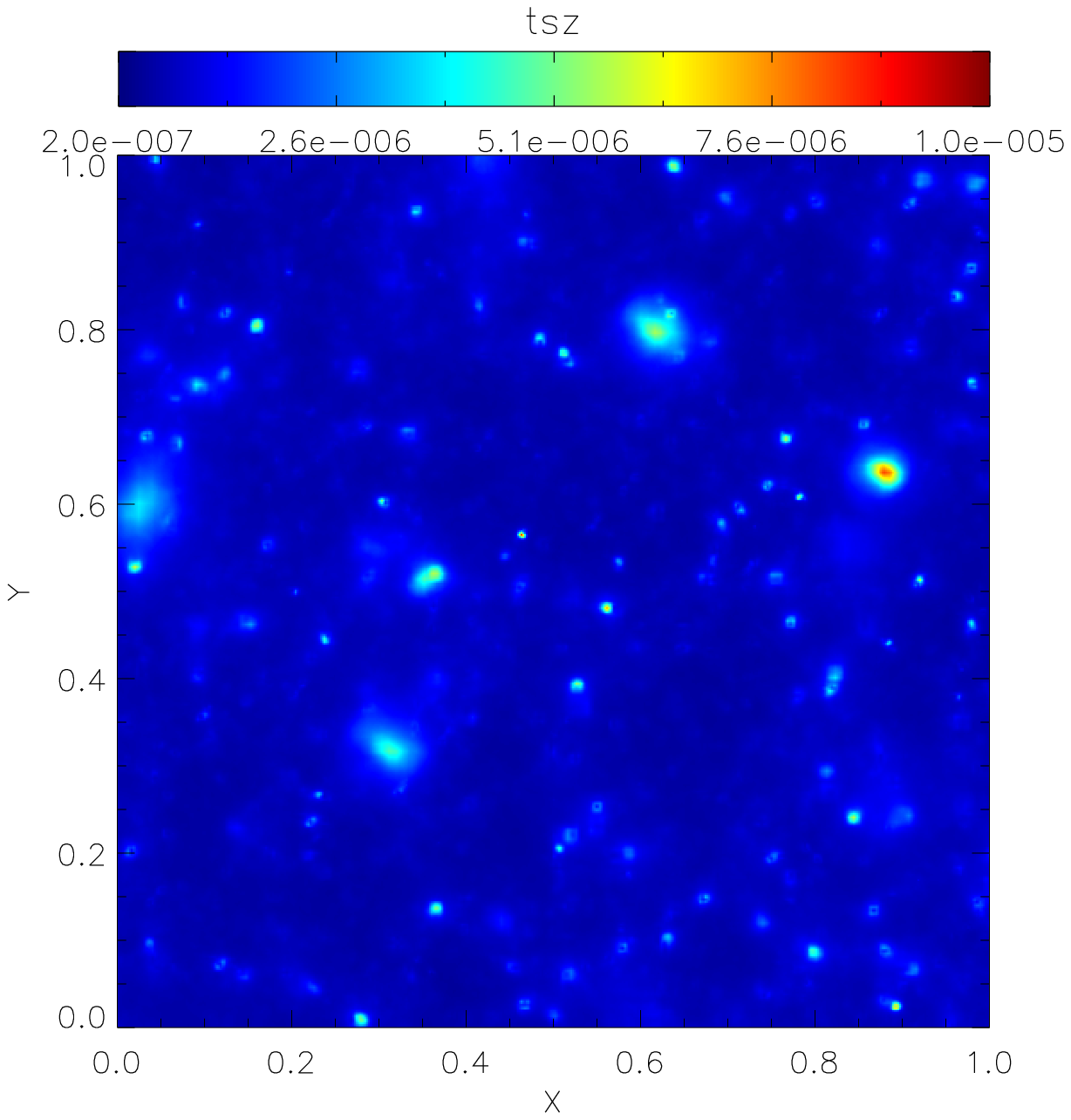}
\end{center}
\vspace{-8mm} \caption{Maps of the kSZ effect (left) and the tSZ effect(right),showing the $b$, $y$ parameter in 
a $15' \times 15'$ patch, respectively.}
\end{figure}

We produce two-dimensional  maps of the kSZ and tSZ effects using a conventional
method,  e.g., refer to Gnedin \& Jaffe(2001), or Zhang et al(2002).
During the cosmological simulation,  two-dimensional  projections of the kSZ and
tSZ effects through the whole simulation box
along the x-, y- and z-directions are stored as sectional maps at different redshifts since $z=6$,
as both the kSZ and tSZ effects
are dominated by contributions from $z=6$ to $z=0$(Roncarelli et al. 2007). The
redshift intervals between the successive outputs are given by the light-crossing
time through the box.  We have 240 and 60 outputs for the 25 $h^{-1}$ Mpc
and 100 $h^{-1}$ Mpc simulation run respectively. We then stack sectional maps,
one at each outstored redshift,
to make a single kSZ or tSZ map. Each sectional map is randomly selected from one of the three
box projections along different axes direction at the corresponding redshift.
To reduce the artificial replication effect in the simulated periodic universe
(Gnedin \& Jaffe 2001), we randomly
place the center of the selected 2-D sectional map, and then rotate and flip it
around any of the six box edge.
We compose $50$ maps of  the kSZ and tSZ effects respectively in size of
$15'$ on a side, corresponding to the angle the box extended at z=6, with pixel
resolution $512 \times 512$, and $1024 \times 1024$ for the 25 $h^{-1}$ Mpc run.
We also produce 50 maps of the kSZ and tSZ effects in size of $1^{\circ}$ respectively
with pixel resolution $1024\times 1024$ for the 100 $h^{-1}$ Mpc run. Figure 1
demonstrates an example
of the kSZ and tSZ maps obtained from the 25 $h^{-1}$ Mpc run. Although the kSZ
effect is smaller than the tSZ
effect in magnitude at center region of dense halos, it is comparable to, and
even larger than the tSZ effect in their outskirts.

{\it 2. Power spectrum.} The power spectra of
the kSZ(b) and tSZ(y) effects are shown in the left
panel of Figure 2,  which are obtained by averaging over 50 maps
of the 25 $h^{-1}$ Mpc run respectively under the small-angle approximation.
The power spectrum of the kSZ
effect grows rapidly when the multipole parameter $l>10^4$, and exceeds
the tSZ power spectrum at around $l\sim 2\times 10^{4}$, corresponding to
the angular scales $\leq 0'.5$.  This behavior is basically consistent with
result given by the lognormal model of the IGM (Astro-Barandela et al.
2008) and similar to those in Zhang et al.(2004) and Roncarelli et
al.(2007). We also present the power spectrum of the kSZ and tSZ maps
integrated from $z=6$ to 4, 2, and 1 in Fig. 2.  It shows that the kSZ power
spectrum is mostly contributed by the electron motion in the redshift range
of $z\sim 1 -2$. This redshift dependence can be interpreted in term of the development history of
the IGM turbulence in the expanding universe recognized recently by Zhu et al.
(2010). Turbulence and vorticity, i.e., the curl part of velocity,
 in the IGM remains weak till $z\sim 4$, and thereby the level of the
power spectrum of the kSZ effect would be low at $z>4$. After $z\sim 1$, both the covered scale
range and intensity of turbulence and vorticity will grow continuously. However,
the number density of electrons $n_e$ in the outskirts of collapsed halos decreases with the cosmic expansion,
the integrated kSZ effect gains most of its weight in the redshift range $z \sim 2-1$.
In the same redshift range, the velocity field of IGM is in the state of fully developed turbulence
between $0.2 h^{-1}$ Mpc and $0.8 h^{-1}$ Mpc, corresponding to the angle scale range of
$0'.2 - 1'.0$, i.e.,$10^4 < l < 5\times 10^4$, within which the peak of the kSZ power spectrum
appears.

\begin{figure}[htb]
\begin{center}
\includegraphics[width=8.0cm]{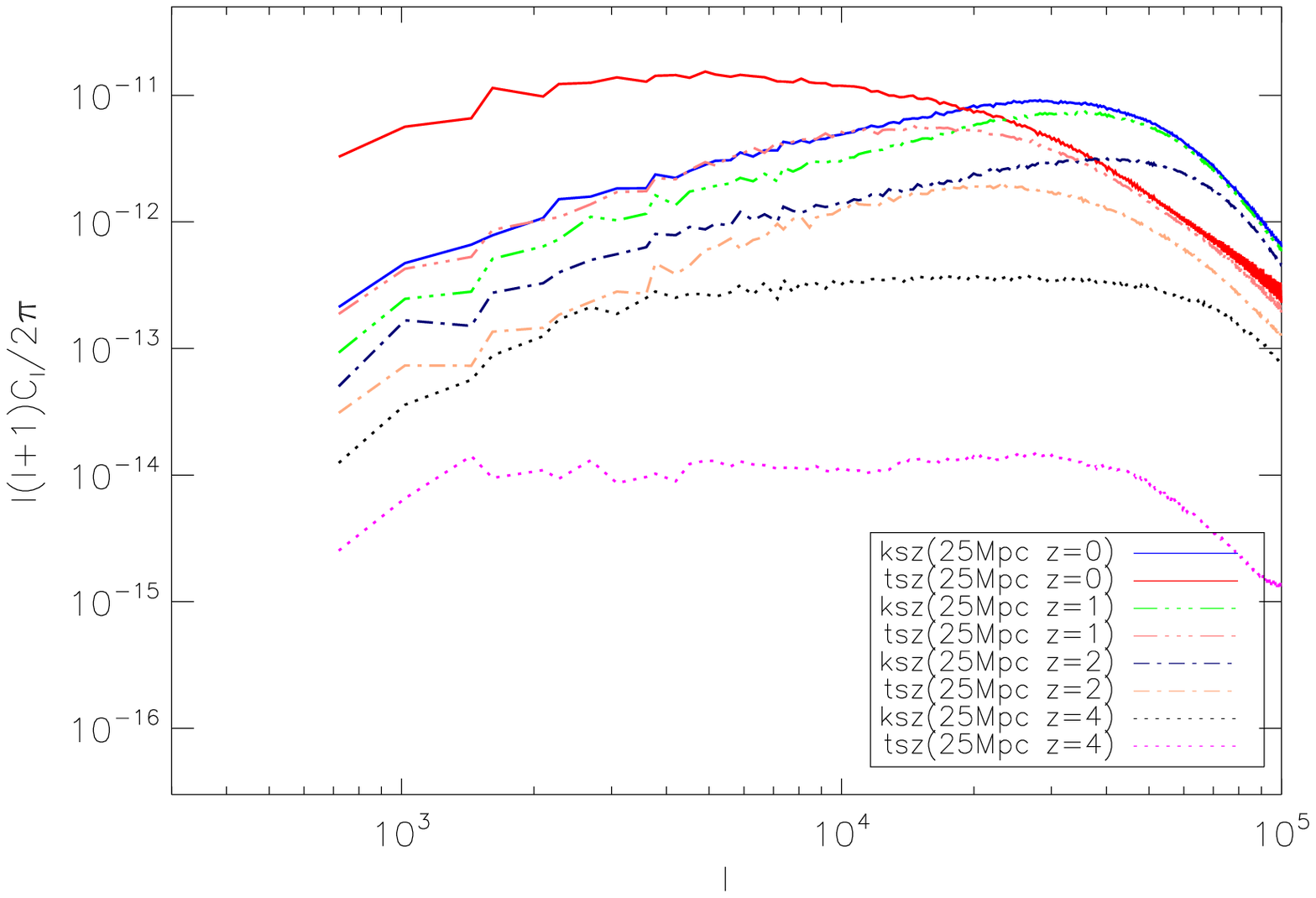}
\includegraphics[width=8.0cm]{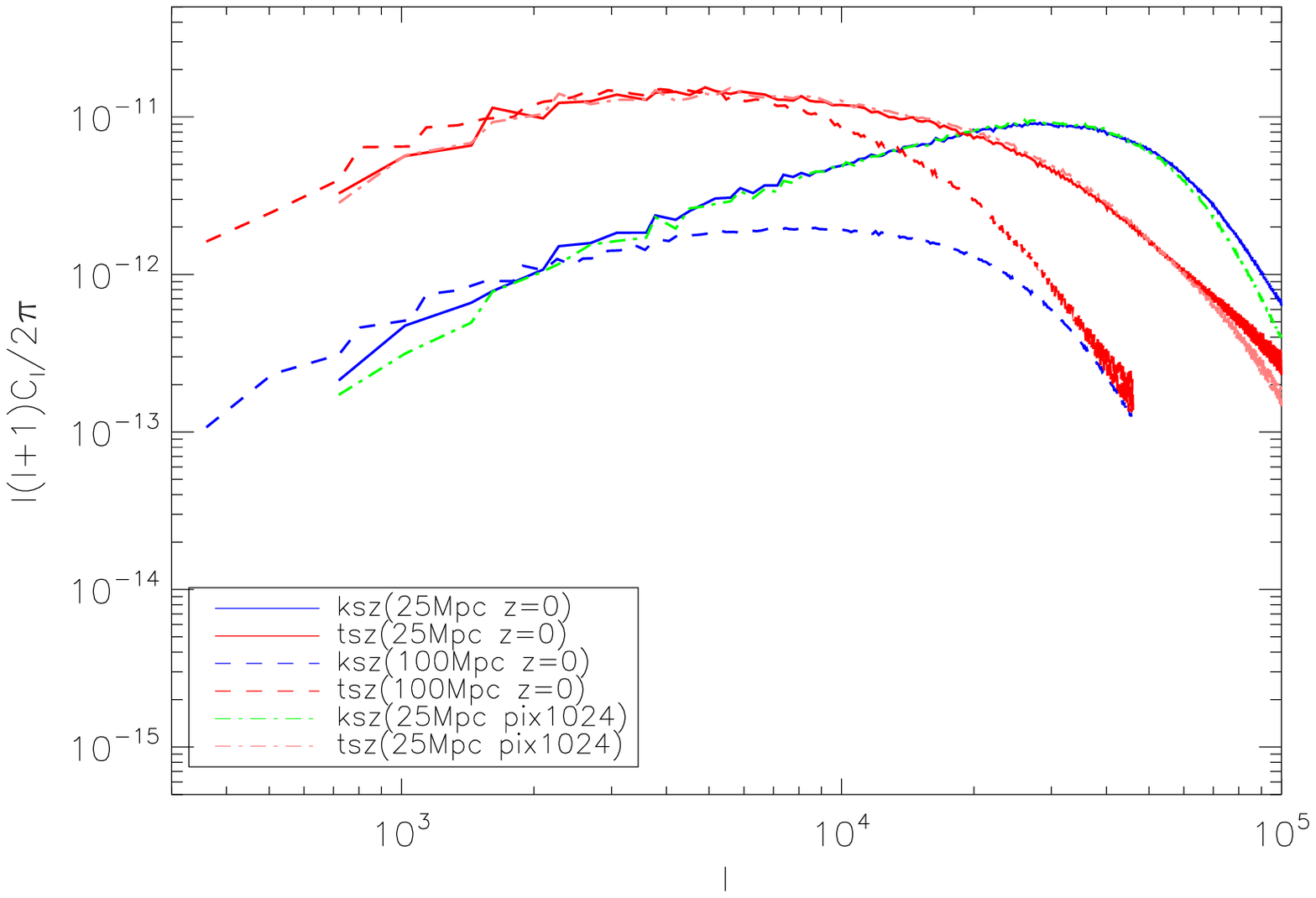}
\end{center}\vspace{-7mm} \caption{Left: the power spectrum of the kSZ (solid
blue) and tSZ  (solid red) effects integrated from
$z=6$ to 0 and averaged over 50 samples produced by 25$h^{-1}$ simulation
 with pixel resolution $512^{2}$; also the results that integrated from $z=6$ to
$z=4$, 2, 1 are shown .Right: Comparison to
the power spectrum integrated from $z=6$ to $z=0$ for samples of 25 $h^{-1}$ Mpc
simulation with pixel resolution $1024\times 1024$(dash), and 100 $h^{-1}$ Mpc
simulation with pixel resolution $1024\times 1024$ (dash dotted).}
\end{figure}

To check the numerical convergence,  we compare the power spectrum of the kSZ and
tSZ effects generated with different pixel resolutions or simulation box size in
the right panel of Figure 2.  Obviously, once the simulation resolution is given, the
resolution of SZ maps has a small impact on the result. The 25 $h^{-1}$
Mpc run may underestimate the power spectrum of the kSZ and tSZ effects due to the
lack of large scale perturbations. However, we find that this
effect is mildly weak on scales $l \geq 10^{3}$ . Although the lack of
perturbations large than the box size may affect the
power spectrum on scales $l < 10^{3}$, the primary temperature
fluctuations will dominate over the tSZ and kSZ effects(Springel et al. 2001;
Zhang et al. 2004; Roncarelliet al. 2007). The statistical features in that scale range
are out the scope of this letter. On the side of small scales, $l\geq 10^4$, the power spectrum
of the 100 $h^{-1}$Mpc simulation run falls rapidly, and is significantly lower
than the 25 $h^{-1}$Mpc run. This feature has been addressed in previous works (Gnedin \& Jaffe 2001;
Zhang et al. 2004), which should result from the poor resolution of the 100 $h^{-1}$Mpc simulation.

\section{Intermittence of the tSZ and tSZ maps}

{\it 1. Structure functions}. To identify the intermittence of the kSZ maps, we use the
dimensionless structure functions defined by
\begin{equation}
S_p(\theta)=\frac{\langle (\delta b_{\theta})^p\rangle} {\langle
(\delta b_{\theta})^2\rangle^{p/2}},
\end{equation}
where $b_{\theta}\equiv b(\hat{\bf n})- b(\hat {\bf n'})$. $\theta$ is
the angle between two directions $\hat{\bf n}$ and $\hat{\bf n'}$. For a Gaussian
field, $S_p(\theta)$ is independent of $\theta$ and equals to
$(p-1)!!$, here $p$ is even integer (Pando et al 2002). The structure functions
of the tSZ maps can be calculated by replacing $b(\hat{\bf n})$ with $y(\hat{\bf
n})$ in eq(2).

The intermittence of the velocity field ${\bf v}$ in fully
developed turbulence is characterized by $\langle (\delta
v_{r})^p\rangle \sim r^{\zeta(p)}$, where the intermittent exponents
$\zeta(p)$ are given by the SL scaling law(She \& Leveque 1994, She et al.
2001). If the kSZ maps contains signatures of the intermittence
caused by the fully developed turbulence in the IGM,
the structure functions should follow a power law
\begin{equation}
S_p(\theta)\propto \theta^{-\xi(p)}.
\end{equation}
According to the SL scaling law,  the intermittent exponents are given by
\begin{equation}
\xi(p)=(p/2)\zeta(2)-\zeta(p)=\alpha(1+\beta)[p - 2(1-\beta^p)/(1-\beta^2)],
\end{equation}
where $\alpha$ and $\beta$ are free parameters, and
$\xi(2)=0$.  The non-Gaussianity is typically measured by parameter $\beta$
(Liu \& Fang 2008). In a Gaussian field, $\beta=1$, while in a intermittent field
$\beta<1$. The basic features of the
intermittent exponents $\xi(p)$ are  1.)
when $p$ is small, $\xi(p)$ is a nonlinear function of $p$, and has
$d^2\xi(p)/dp^2 >0$; 2.) when $p$ is large, $\xi(p)$ has
to be a straight line with slope $\alpha(1+\beta)$ as $\beta<1$;

The non-Gaussianity of the kSZ maps is affected by two major factors. First, the kSZ effect
is given by an integral of the momentum density of the electrons, $n_e{\bf v}$
along the line of sight.
The non-Gaussian behavior of $n_e{\bf v}$  will be developed significantly
by intermittence either in the
velocity field ${\bf v}$  or the electrons density field $n_e$.  If both of these
two fields are intermittent,
the multiplication will lead to enhancement of the non-Gaussianity in the momentum
density field.
Second, the maps of $b({\bf n})$ are given by the summation of the
momentum field $n_e{\bf v}$ in successive redshifts intervals, therefore the
non-Gaussianity might be reduced according to the central limit theorem. However,
the statistical behaviors governed by the central limit theorem will be remarkably retarded by the
development history of the IGM turbulence. As mentioned in \S 2, the integrated kSZ effect
will be dominated by such turbulent flows fully developed in the range $z\sim 1 -2$.  Moreover,
for intermittent fields in which large deviation events always occur,
the central limit theorem acts slowly (Jamkhedkar et al 2003, Fujisaka \& Inoue 1987;
Frisch 1996). In theory,
the kSZ maps generated by eq.(1) may still contain the
signatures of the non-Gaussian statistical properties of turbulence.
The case with the tSZ effect is different as its non-Gaussianity is mainly given
by the electrons' density and temperature field.

{\it 3. Intermittent exponents.} Figure 3 presents the structure
functions of the kSZ maps in the
 angular range $0'.05 - 1'.5$. As a comparison, we also plot
the structure functions of the tSZ and total SZ
effects, the latter is the superposition of the kSZ and tSZ effects.

A remarkable feature displayed in Figure 3 is that the structure functions
of the kSZ, tSZ and total SZ maps are in agreement with the power law
in eq.(3). It indicates that all the maps are essentially non-Gaussian.
In addition, the correlation between $\ln S_p(\theta)$ and $\ln \theta$ of the kSZ effect
is much stronger than that of the tSZ effect, suggesting that the
non-Gaussianity of the kSZ maps is much significant.
\begin{figure}[htb]
\begin{center}
\includegraphics[width=5.2cm]{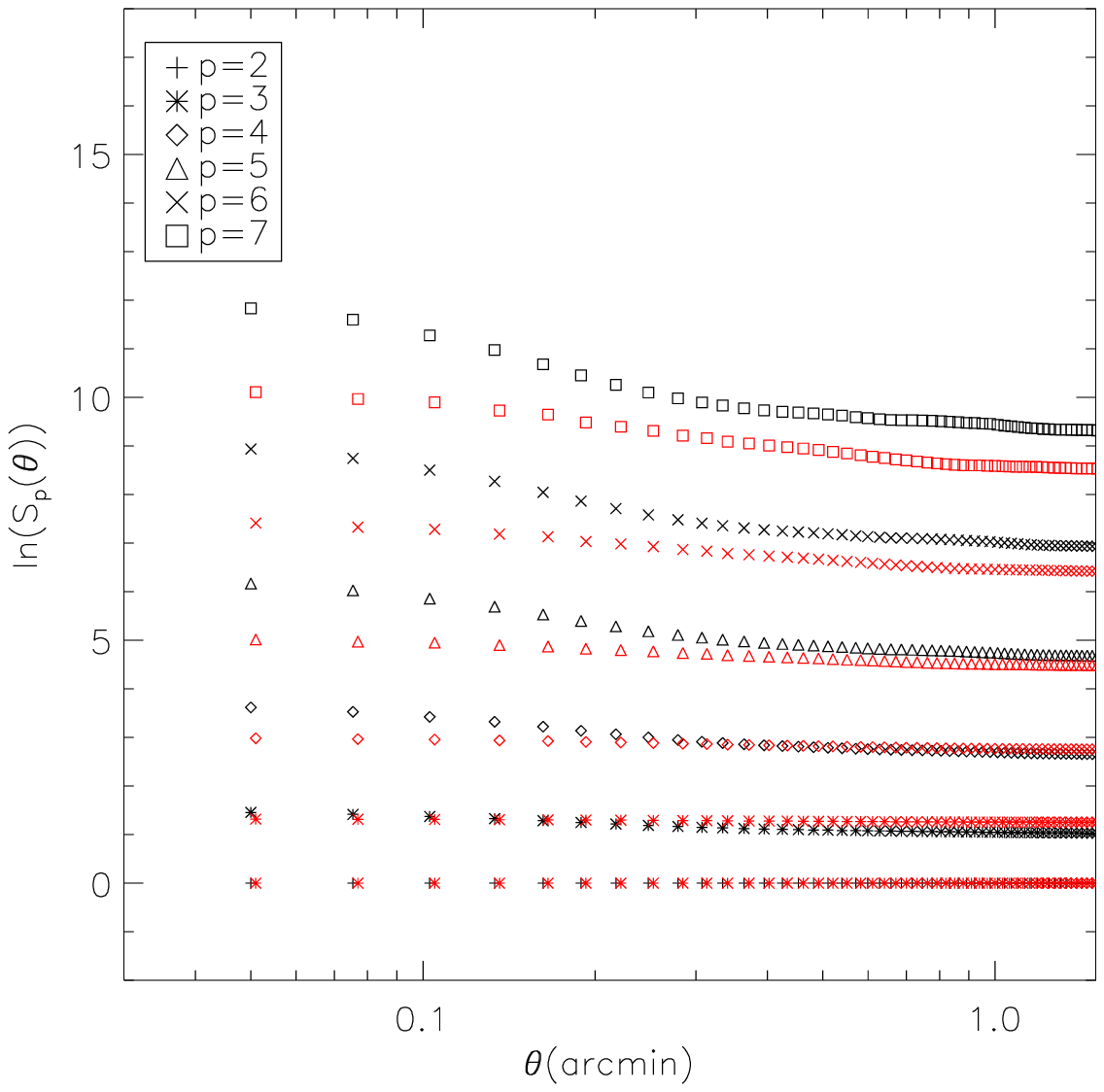}
\includegraphics[width=5.2cm]{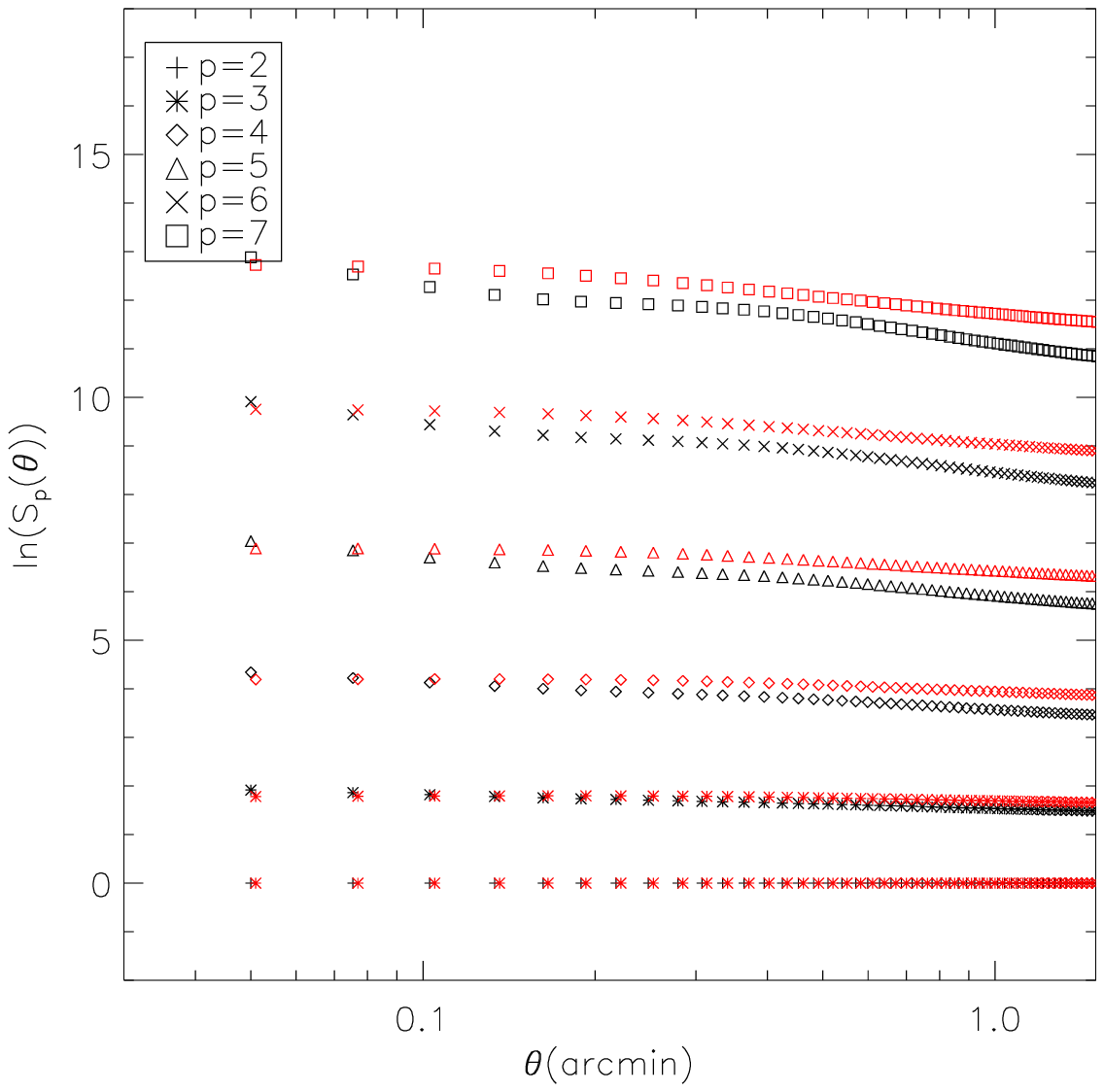}
\includegraphics[width=5.2cm]{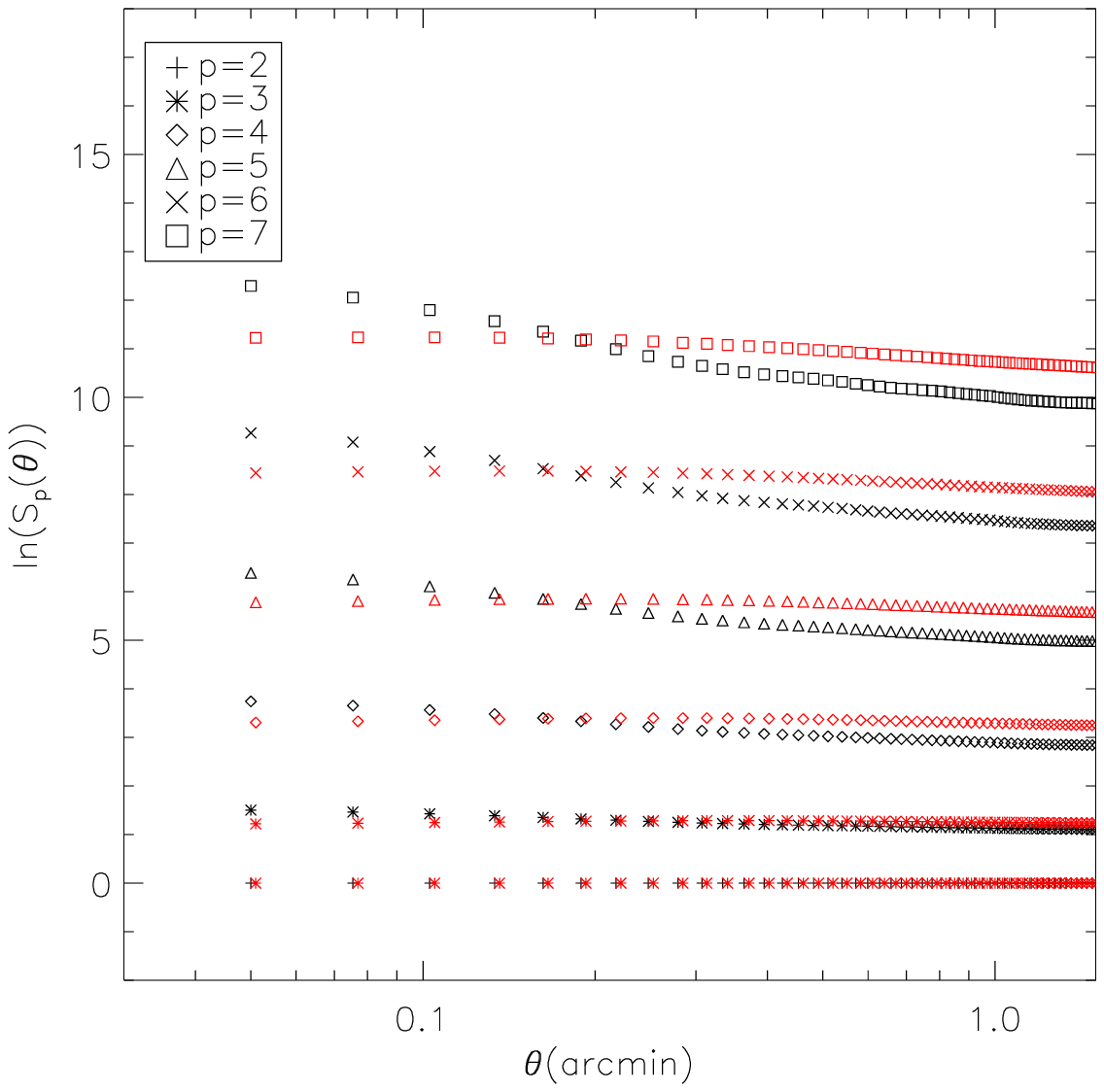}
\end{center}
\vspace{-7mm} \caption{ $\log S_p(\theta)$ vs. $\theta$ of the maps
of the kSZ (left), tSZ (middle) and total SZ (kSZ + tSZ) (right) effects.
 They are averaged over 50 samples. Black and red curves represent
the samples of 25 $h^{-1}$ and 100 $h^{-1}$ Mpc respectively.}
\end{figure}

Figure 3 shows that on scales $\theta > 0.5$ arcmins
the logarithmic structure function
$\log S_p(\theta)$ of the kSZ effect obtained in the  25 $h^{-1}$ and 100 $h^{-1}$ Mpc
simulations
remain the same till $p=5$, and the deviations on 6th
and 7th orders are visible, but still less than 10\%. On small scales $\theta<0.5$ arcmins,
the logarithmic structure functions of the kSZ  effect in the 100 $h^{-1}$ Mpc run are
clearly lower than
the 25 $h^{-1}$  Mpc run.  However, for the tSZ effect, we do not see any significant
difference between these two simulations in overall angular range $0'.05 - 1'.5$.  This feature is most
likely caused by
the velocity field having stronger dependence on the simulation
resolution than the temperature field, which has been actually recognized by the power spectrum comparison.
It indicates that the kSZ is more sensitive to the non-Gaussianity of the turbulence, namely,
the non-Gaussianity of the total SZ maps should be largely attributed to the kSZ effect.

The intermittent exponents $\xi(p)$ as a function of the order $p$ are plotted in
Figure 4. For the samples produced in the 25 $h^{-1}$ Mpc simulation,
the intermittent exponents $\xi(p)$  display a tight linear relation with the order parameter $p$
in the kSZ maps, but is relatively weak for the tSZ maps. The intermittent exponents $\xi(p)$
can be best fitted as a function of $p$ by eq.(4) with
$\alpha =0.18,\ \beta=0.25$, $\alpha =0.08,\ \beta=0.7$ and $\alpha =0.11,\ \beta=0.41$ for the
kSZ, tSZ and total SZ effects, respectively.

In comparison, the right panel of Figure 4 presents the intermittent exponents
$\xi(p)$ measured in the 100 $h^{-1}$ Mpc simulation. Clearly, they have a weaker
dependence on $p$.  The least-square fitting gives
$\beta=0.85$, $\beta=0.95$ and $\beta=0.92$ for the
kSZ, tSZ and total SZ effects respectively.
It implies that, the non-Gaussianity characterized by the parameter $\beta$ of samples produced by
 100 $h^{-1}$ Mpc simulation is much weaker than the 25 $h^{-1}$ Mpc
 simulation.  It can be concluded that the nonlinear development of
the velocity field at small scales can produce stronger
non-Gaussianity.

\begin{figure}[htb]
\begin{center}
\includegraphics[width=8.0cm]{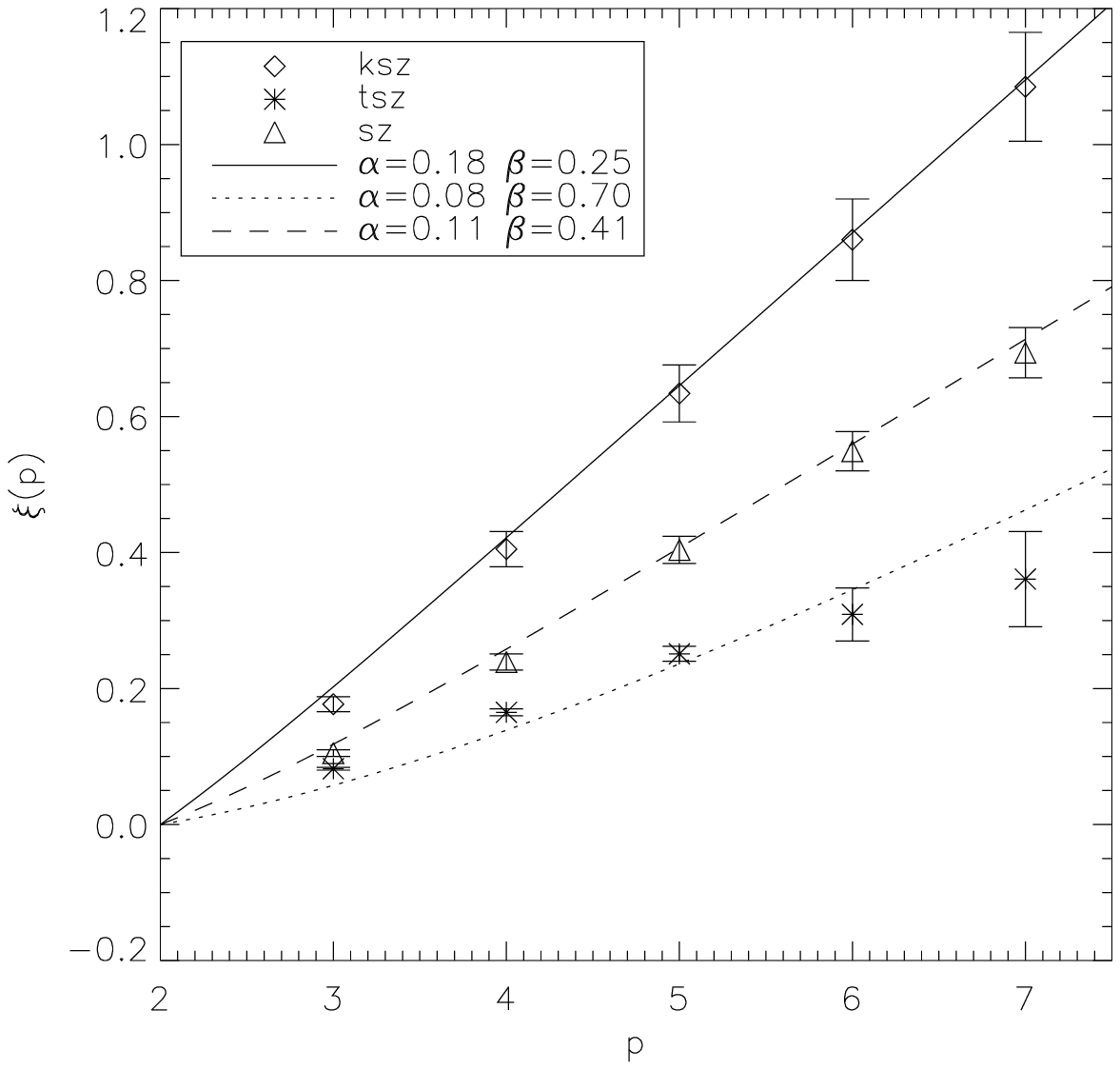}
\includegraphics[width=8.0cm]{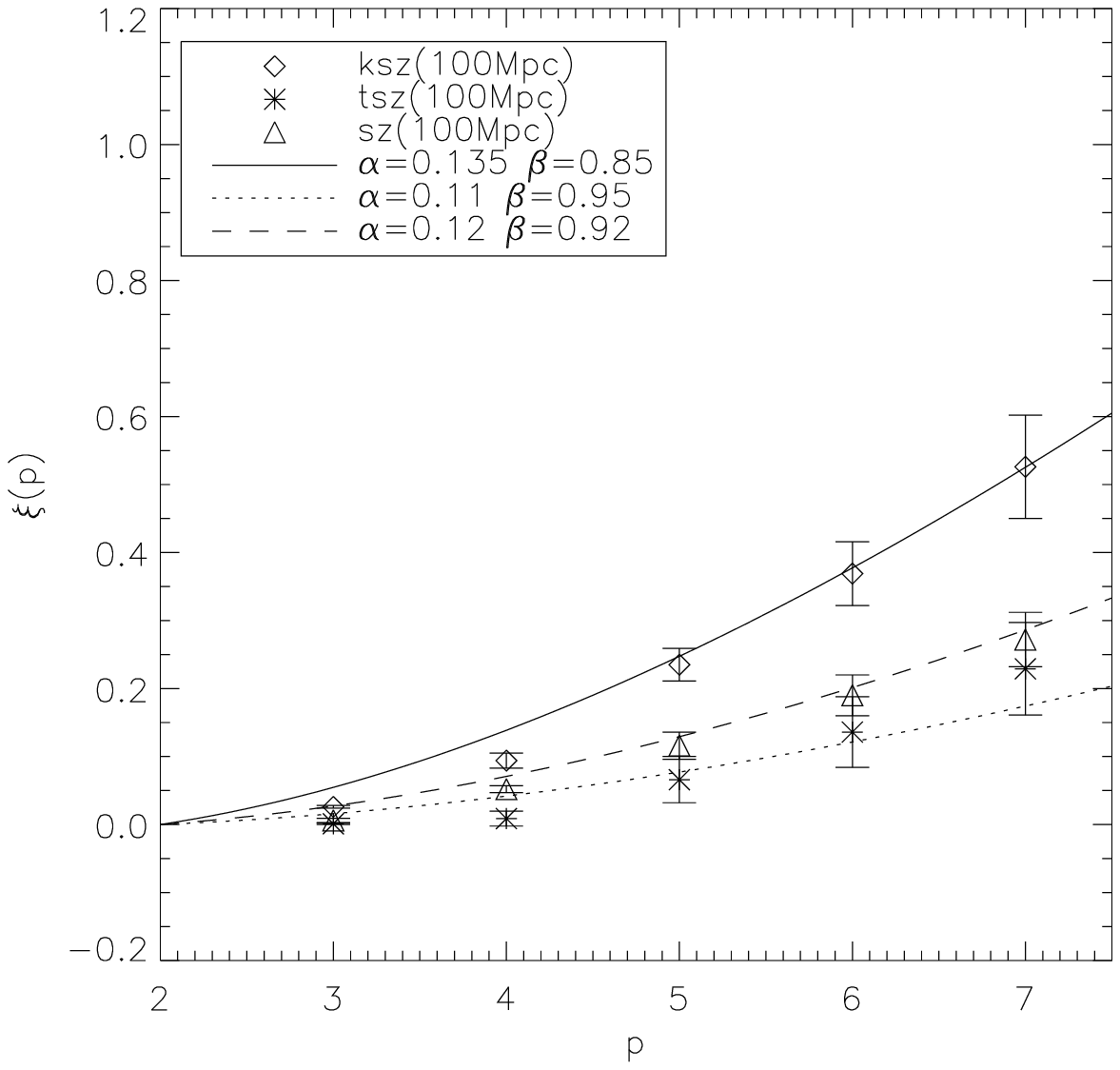}
\end{center}
\vspace{-7mm} \caption{The intermittent exponents of the maps of kSZ,
tSZ and (kSZ+tSZ)
in 25 Mpc(left) and 100 Mpc(right) simulation box.}
\end{figure}

\section{Discussions and Conclusions}

With cosmological hydrodynamical simulation in the concordance
$\Lambda$CDM model, the turbulent cosmic baryon fluid is found to
yield intermittence in the second order CMB temperature fluctuation due to
 the kSZ effect, which can be described in term of the structure functions.
The structure functions can be applied to identify the types of the
non-Gaussianity (Pando et al 2002). We demonstrate that the structure
functions of the kSZ maps possess the typical behavior of fully
developed turbulence, i.e. the intermittent exponents follow a law similar to
the She-Leveque universal scaling.  Although the weakly intermittent tSZ effect will be
dominant over the kSZ effect in the center regions of virialized halos, the total SZ
effect still displays significant intermittence. Therefore, it is
expected that the intermittence of the kSZ effect would remain
detectable on angular scales $0'.05\sim 1'.0$ in noisy maps,
if the noises from foreground and others sources are not stronger
than the tSZ effect.

Star formation and its feedback on the IGM has not been included in the
simulation. Even though the SZ effect is strong in
massive collapsed objects with hot gas, the intermittent exponents would not be
dominated by the hot collapsed objects, because $S_p$ is given by the ratio of
the fluctuations of the SZ effect [eq.(2)].  Accordingly, it is concluded
that the turbulent behavior of the IGM may be detectable by
high quality observation of CMB temperature fluctuations on angular scales $\leq 1'.0$,
i.e., the multipole parameters $l\geq  2\times 10^4$, which,
however, is below the detection limit of current observational projects.

\begin{acknowledgements}

We thanks Dr.Priya Jamkhedkar for her helps. WSZ acknowledges the
support of the International Center for Relativistic Center Network
(ICRAnet). LLF acknowledges the support from the NSFC under grants
of Nos.10633040, 10725314 and the Ministry of Science \& Technology
of China through 973 grant of No. 2007CB815402.

\end{acknowledgements}

\end{document}